\documentclass[pre, showpacs, nofootinbib, floatfix]{revtex4}
\usepackage{latexsym}
\usepackage{amsmath}
\usepackage[dvips]{graphicx}

\begin {document}
\title{Transport on randomly evolving trees}
\author{L. \surname{P\'al}}

%\email[Electronic address:]{lpal@rmki.kfki.hu}

\affiliation{KFKI Atomic Energy Research Institute, H-1525
Budapest 114, POB 49, Hungary}

\begin{abstract}
The time process of transport on randomly evolving trees is
investigated. By introducing the notions of living and dead nodes, a
model of random tree evolution is constructed which describes the
spreading in time of objects corresponding to nodes. It is assumed
that at $t=0$ the tree consists of a single living node (root), from
which the evolution may begin. At a certain time instant $\tau \geq
0$, the root produces $\nu \geq 0$ living nodes connected by lines
to the root which becomes dead at the moment of the offspring
production. In the evolution process each of the new living nodes
evolves further like a root independently of the others. By using
the methods of the age-dependent branching processes we derive the
joint distribution function of the numbers of living and dead nodes,
and determine the correlation between these node numbers as a
function of time. It is proved that the correlation function
converges to $\sqrt{3}/2$ independently of the distributions of
$\nu$ and $\tau$ when $q_{1} \rightarrow 1$ and $t \rightarrow
\infty$. Also analyzed are the stochastic properties of the
end-nodes; and the correlation between the numbers of living and
dead end-nodes is shown to change its character suddenly at the very
beginning of the evolution process. The survival probability of
random trees is investigated and expressions are derived for this
probability.


\end{abstract}

\maketitle

\section{Introduction}

Recently there has been a growing interest in randomly evolving
graphs since a large number of practically important problems can be
analyzed in this way. In the last few years so many papers have been
presented in this field that any list of references would be far
from complete by any standards. It is fortunate that two outstanding
review papers \cite{barabasi01}, \cite{dorogovtsev01} have been
published in 2002. Thus the author of the present paper does not
feel obliged to cite the very large amount of important references.
In addition, one must mention the excellent book by Dorogovtsev and
Mendes \cite{dorogovtsev02}, which gives a very interesting summary
of the actual problems of the growth and structure of random
networks.

It has been long known that the random evolution of a
tree~\footnote{A randomly evolving tree is a connected graph
containing no cycles, and growing from a single node (root)
according to well-defined rules.} corresponds to a Galton-Watson
process \cite{harris63} which describes the formation of a
population which starts at $t=0$ with one individual and in which at
discrete time moments $t=1, 2, \ldots$ each individual produces a
random number of offspring with the same distribution independently
of the others. There are many interesting papers on this subject,
statements of which are analyzed and generalized in Ref.
\cite{lyons02}. The purpose of this paper is to study the random
tree evolution with continuous time parameter with the hope that
this may help to a better understanding of the nature of the
age-dependent branching processes which play a significant role in
many fields of physical and biological sciences \cite{janossy52,
reid54, lpal58, kendall60, bell63, pazsit99, lpal03, williams04}.
The general theory of age-dependent branching processes is
excellently summarized in Ref. \cite{harris63} and in Ref.
\cite{sevast'yanov71}, therefore we decided to focus only on the
correlation and survival properties of the randomly evolving trees.

The paper is organized as follows. In Sec. II the exact description
of the random process is given, while in Sec. III the derivations of
the basic equations for the generating functions can be found. The
properties of average characteristics (expectation values,
variances, and correlation functions) are discussed in Sec. IV.
Survival probability of trees is determined in Sec. V, and finally,
the conclusions are summarized in Sec. VI. It is to mention that for
more detailed analysis of the problem see \cite{lpal02}.

\section{Description of the random process}

We will study the random evolution of trees which consist of living
and dead nodes as well as lines connecting the nodes. A living node
is unstable, and after a certain time $\tau$ may become dead. The
lifetime $\tau$ is a random variable, and ${\mathcal P}\{\tau \in
(t, t+dt)\} = dT(t)$ denotes the probability of finding $\tau$ in
the time interval $(t,t+dt)$. A dead node is inactive, it cannot
change its state, and is unable to interact with the other nodes.

The evolution process can be described as follows. Let us suppose
that at time instance $t=0$ the tree consists of a single living
node called the root and let ${\mathcal S}_{0}$ denote this state of
the tree. At the end of its life the root creates $\nu =0, 1, 2,
\ldots\;$ new living nodes and immediately after that it becomes
dead. The number $\nu$ of the offspring  is a random variable, and
the probabilities ${\mathcal P}\{\nu=k\} = f_{k}, \;\; k = 0, 1, 2,
\ldots$ are assumed to be known. The new living nodes are promptly
connected to the dead one and each of them can evolve further like a
root independently of the others. This means that the tree from its
initial state ${\mathcal S}_{0}$ may evolve further by two mutually
exclusive ways, in which the root either
\begin{itemize}
\item  survives, and thus the tree remains in its
initial state during the whole time interval $(0, t)$, or \item the
root does die in an infinitesimally small subinterval $\;(t', t' +
dt') \in (0, t)\;$ and produces $j \geq 0$ new living nodes. Each of
these nodes will evolve further in the remaining time interval
$(t',t)$ independently of the others like the root.
\end{itemize}

\begin{figure}[ht]
\protect \centering{
\includegraphics[height=7.5cm, width=7.5cm] {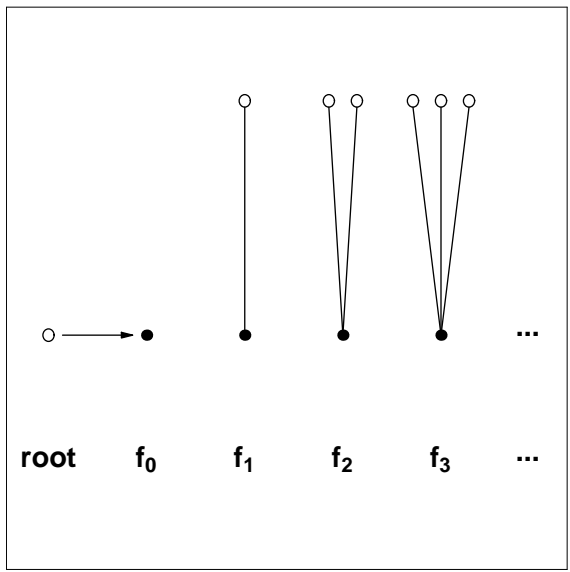}}\protect
\caption{\label{fig1} \footnotesize{First step of the tree
evolution. The new state arises from the initial state which
consists of a single root. Four of the possible new states are
illustrated in the figure. Living nodes are denoted by white circles
and the dead ones by black circles. $f_{0}, f_{1}, f_{2}, f_{3},
\ldots\;$ are the probabilities of producing $0, 1, 2, 3, \ldots\;$
nodes by one dying root.}}
\end{figure}

Figure. \ref{fig1} shows the initial state of a tree consisting of a
single root and illustrates four of the possible states which may be
produced by the dying root. The probabilities of producing $0, 1, 2,
3, \ldots$ living nodes are denoted by $f_{0}, f_{1}, f_{2}, f_{3},
\ldots\;\;$.

\begin{figure}[ht]
\protect \centering{
\includegraphics[height=8cm, width=8cm] {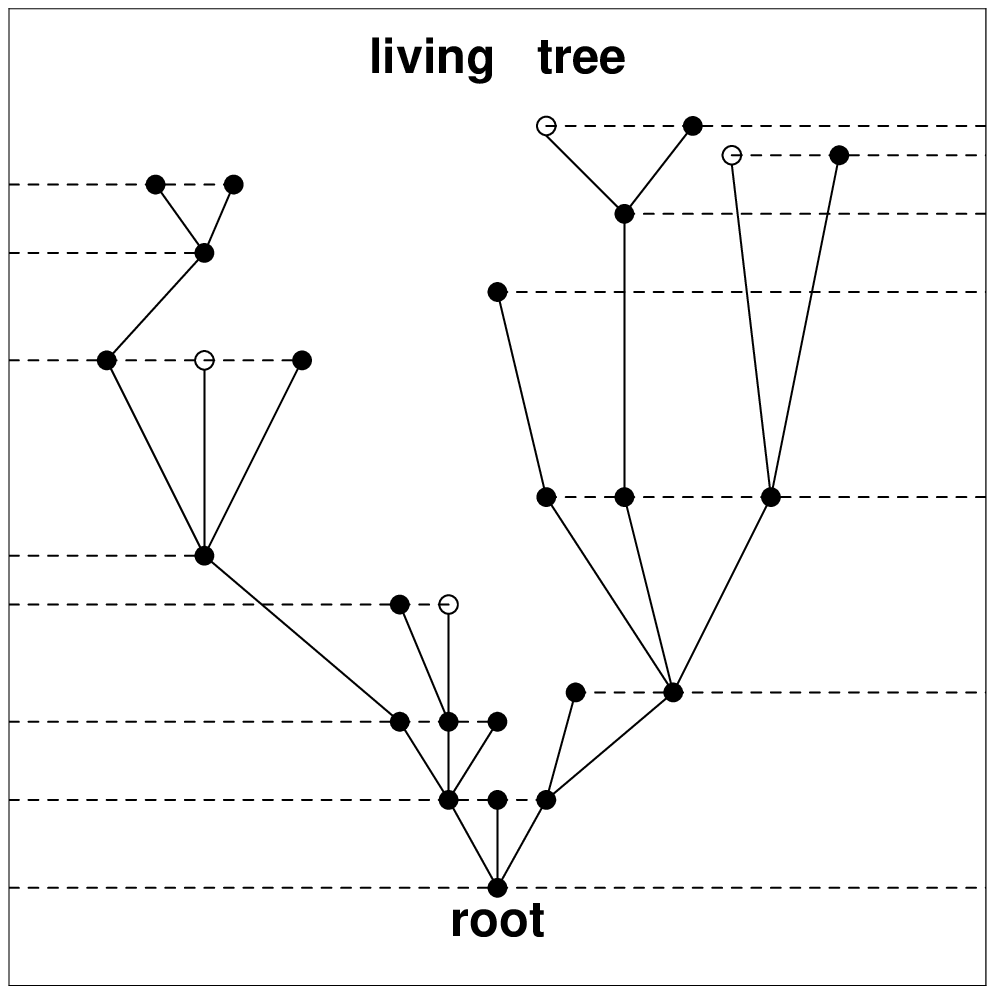}}\protect
\caption{\label{fig2}\footnotesize{One of the realizations of a
random tree at the time moment $t > 0$. The nodes denoted by white
circles are capable of further evolution, while those denoted by
black cannot evolve and further the horizontal dashed lines
indicate the time instances at which nodes were produced.}}
\end{figure}

Figure. \ref{fig2} illustrates one of the realizations of a random
tree at the moment $t > 0$. In Fig. \ref{fig2} there are three white
circles denoting living nodes which are capable of further
evolution,  and a large number of black circles denoting dead nodes,
which are unable to produce new living nodes. It is to mention that
each offspring is connected by a line to its dead precursor. If a
dead node has $k$ outgoing lines it is called node of $k$th
out-degree.

In order to simplify our consideration we will assume that the
distribution function of the lifetime $\tau$ of living nodes is
exponential, i.e.,
\begin{equation} \label{1}
dT(t) = e^{-Qt}\;Q\;dt,
\end{equation}
where $1/Q$ is the expectation value of the lifetime. In this case
the evolution becomes a Markovian  process. It is assumed that the
probabilities
\begin{equation} \label{2}
{\mathcal P}\{\nu = k\} = f_{k}, \;\;\;\;\;\; k = 0, 1, 2, \ldots
\end{equation}
are the same for all living nodes. By introducing the generating
function
\begin{equation} \label{3}
q(z) = \sum_{k=0}^{\infty} f_{k}\;z^{k}
\end{equation}
we can define the factorial moments of $\nu$ in the following way:
\[ q_r = \left[\frac{d^{r}q(z)}{dz^{r}}\right]_{z=1} \;\;\;\;\;\; r
= 1, 2, \ldots\;\;. \] For the expectation value and the variance
of $\nu$ we have
\begin{equation} \label{4}
{\bf E}\{\nu\} = q_1 \;\;\;\;\;\; \mbox{and} \;\;\;\;\;\; {\bf
D}^{2}\{\nu\} = q_2 + q_1 - q_1^{2}.
\end{equation}
For the sake of later use we cite the relation
\begin{equation} \label{5}
{\bf E}\{(\nu-1)^{2}\} = {\bf D}^{2}\{\nu\} + (1-q_1)^{2}.
\end{equation}

In many cases we do not have to know the exact form of probabilities
$f_{k}, \; k \in {\mathcal Z}$, it is enough to know only the first
and second factorial moments of $\nu$. However, if we wish to obtain
results which are, in some sense, exact then we will assume that
$f_{k}=0, \;\; \forall k>2$, i.e. we will use a quadratic generating
function
\begin{equation} \label{5a}
q(z) = f_{0} + f_{1} z + f_{2} z^{2} = 1 + q_{1} (z-1) +
\frac{1}{2} q_{2} (z-1)^2.
\end{equation}
It seems to be useful to cite the following trivial relations:
\[ f_{0} = 1 - q_{1} + \frac{1}{2}\;q_{2}, \;\;\;\;
f_{1} = q_{1} - q_{2},  \;\;\;\;  f_{2} = \frac{1}{2}\;q_{2}, \]
which follow from the equations $f_{0} + f_{1} + f_{2}  =  1,
\;\;\;\; f_{1} + 2f_{2} = q_{1},\;$ and $2\;f_{2} = q_{2}$. Since
$f_{0}, f_{1}$, and  $f_{2}$ are non-negative real numbers smaller
than $1$, and their sum is equal to $1$, the possible values of
$q_{1}$ and $q_{2}$ are restricted.

As a source of simplification, sometimes we use the geometric
distribution
\[{\mathcal P}\{\nu=k\} = \frac{1}{1+q_{1}}\;
\left(\frac{q_{1}}{1+q_{1}}\right)^{k},  \] which is defined by
the single parameter $q_{1}$ only. The generating function is
expressed by
\begin{equation} \label{5b}
q_{g}(z) = \frac{1}{1+(1-z)q_{1}},
\end{equation}
and we find immediately that ${\bf E}\{\nu\} = q_{1}$ and ${\bf
D}^{2}\{\nu\} = q_{1}(1+q_{1})$, i.e.,  $q_{2} = 2 q_{1}^{2}$.

In order to characterize the random process of tree evolution we
introduce the random functions $\mu_{\ell}(t)$ and $\mu_{d}(t)$
which are the momentary numbers of the living and dead nodes,
respectively, at time instant $t \geq 0$. It seems to be worthwhile
to define the random function $\mu(t) = \mu_{\ell}(t) + \mu_{d}(t)$
which gives the total number of nodes, i.e., the momentary value of
the tree size at $t \geq 0$.

The nodes can be sorted according to the number of outgoing lines.
Let us denote by $\chi_{\ell}(t,k)$ the number of living, while by
$\chi_{d}(t,k)$ the number of dead nodes with $k$ outgoing lines, at
the time moment $t$. A node not having any outgoing lines is called
an end-node. Obviously, all living nodes are end-nodes, i.e.,
$\chi_{\ell}(t,0) = \mu_{\ell}(t)$ and $\chi_{\ell}(t,k) = 0$, if $k
> 0$. Clearly,
\[ \mu_{d}(t) = \sum_{k=0}^{\infty} \chi_{d}(t,k). \] To simplify the
notation, we introduce the following random functions:
\[ \chi_{\ell}(t) = \chi_{\ell}(t,0) \;\;\;\;\;\; \mbox{and}
\;\;\;\;\;\; \chi_{d}(t) = \chi_{d}(t,0). \] The total number of
end-nodes is given by $\chi(t) = \chi_{\ell}(t) + \chi_{d}(t)$. In
the next section we shall derive the basic equations for the
generating functions of probabilities of the random events
$\{\mu_{\ell}(t) = n_{\ell},\; \mu_{d}(t) = n_{d}\},\;\;\;\{\mu(t) =
n\}\;$ and $\;\{\chi_{\ell}(t) = n_{\ell},\; \chi_{d}(t) = n_{d}\},
\;\;\; \{\chi(t) = n\}$, provided that at $t=0$ the tree was in the
state ${\mathcal S}_{0}$.

\section{Derivation of the basic equations}

In this section we would like to deal with equations for the
generating functions
\begin{equation} \label{5c}
g^{(\ell, d)}(t, z_{\ell}, z_{d}) = {\bf
E}\left\{z_{\ell}^{\mu_{\ell}(t)} \;z_{d}^{\mu_{d}(t)}\right\},
\;\;\;\;\;\; \mbox{and} \;\;\;\;\;\; g(t, z) = {\bf
E}\left\{z^{\mu(t)}\right\},
\end{equation}
as well as for
\begin{equation} \label{5d}
g_{e}^{(\ell,d)}(t, z_{\ell}, z_{d}) = \mathbf{
E}\left\{z_{\ell}^{\chi_{\ell}(t)} \;z_{d}^{\chi_{d}(t)}\right\}
\;\;\;\;\;\; \mbox{and} \;\;\;\;\;\; g_{e}(t, z) =
\mathbf{E}\left\{z^{\chi(t)}\right\}.
\end{equation}
It is seen immediately that the generating functions
$g^{(\ell)}(t, z) = {\bf E}\left\{z^{\mu_{\ell}(t)}\right\}$ and
$g^{(d)}(t, z) = {\bf E}\left\{z^{\mu_{d}(t)}\right\}$ can be
obtained from (\ref{5c}) since
\[ g^{(\ell, d)}(t, z_{\ell}=z, z_{d}=1) = g^{(\ell)}(t, z),
\;\;\;\;\;\; \and \;\;\;\;\;\; g^{(\ell, d)}(t, z_{\ell}=1,
z_{d}=z) = g^{(d)}(t, z), \] and similarly, $g_{e}^{(\ell)}(t, z)$
and $g_{e}^{(d)}(t, z)$ from (\ref{5d}), so it is clear that we
need equations only for the generating functions \[ g^{(\ell,
d)}(t, z_{\ell}, z_{d}),\;\;\;\;\;\;\;g(t, z)\;\;\;\;\;\;
\mbox{and} \;\;\;\;\;\;g_{e}^{(\ell,d)}(t, z_{\ell} z_{d}),
\;\;\;\;\;\; g_{e}(t, z).\] It is evident that these equations
belong to the class of equations corresponding to the
age-dependent branching processes.

\subsection{Joint distribution of the numbers of living and dead
nodes}

We shall define now the probability
\begin{equation} \label{6}
{\mathcal P}\{\mu_{\ell}(t)=n_{\ell}, \mu_{d}(t)=n_{d}|{\mathcal
S}_{0}\} = p^{(\ell,d)}(t, n_{\ell}, n_{d}),
\end{equation}
where $n_{\ell}$ and $n_{d}$ are non-negative integers. It is clear
that $p^{(\ell,d)}(t, n_{\ell}, n_{d})$ gives the probability that
at the time instant $t \geq 0$ the number of living and that of dead
nodes in a randomly evolving tree are equal to $n_{\ell}$ and
$n_{d}$, respectively, provided that at $t=0$ the tree was in the
state ${\mathcal S}_{0}$. By exploiting the rules of random
evolution described in Sec. II and applying the Theorem 7.1 of Ref.
\cite{harris63} it can be shown that
\[ p^{(\ell,d)}(t, n_{\ell}, n_{d}) = e^{-Qt}\;\delta_{n_{\ell},1}\;\delta_{n_{d},0} + \]
\[ Q\;\int_{0}^{t} e^{-Q(t-t')}\left\{f_{0}\;\delta_{n_{\ell},0}\;\delta_{n_{d},1} +
\sum_{k=1}^{\infty}f_{k}\;V^{(\ell,d)}(t', n_{\ell}, n_{d} \vert
k)\right\}\;dt, \] where
\[ V^{(\ell,d)}(t', n_{\ell}, n_{d} \vert k) =
\sum_{n_{\ell,1}+\cdots+n_{\ell,k}=n_{\ell}}\;\sum_{n_{d,1}+\cdots+n_{d,k}=n_{d}-1}
\prod_{j=1}^{k}p^{(\ell,d)}(t', n_{\ell,j}, n_{d,j}). \]
Introducing the probability generating function
\begin{equation} \label{7}
g^{(\ell,d)}(t, z_{\ell}, z_{d}) = \sum_{n_{\ell}=0}^{\infty}\;
\sum_{n_{d}=0}^{\infty} p^{(\ell,d)}(t, n_{\ell},
n_{d})\;z_{\ell}^{n_{\ell}}\;z_{d}^{n_{d}},
\end{equation}
and taking into account the generating function defined by Eq.
(\ref{3}), we obtain the integral equation in the form
\begin{equation} \label{8}
g^{(\ell,d)}(t, z_{\ell}, z_{d})= e^{-Qt}\;z_{\ell} +
z_{d}\;Q\;\int_{0}^{t} e^{-Q(t-t')}\;q\left[g^{(\ell,d)}(t',
z_{\ell}, z_{d})\right]\;dt',
\end{equation}
which according to the Theorem 11.1 of Ref. \cite{harris63} can be
rewritten as a nonlinear differential equation
\begin{equation} \label{9}
\frac{\partial g^{(\ell,d)}(t, z_{\ell}, z_{d})}{\partial t} = -
Q\;g^{(\ell,d)}(t, z_{\ell}, z_{d}) +
z_{d}\;Q\;q\left[g^{(\ell,d)}(t, z_{\ell}, z_{d})\right]
\end{equation}
with initial condition
\[ \lim _{t \downarrow 0}\; g^{(\ell,d)}(t, z_{\ell}, z_{d}) = z_{\ell}. \]

\subsubsection{Remarks}

By using the relations $g^{(\ell, d)}(t, z_{\ell}=z, z_{d}=1) =
g^{(\ell)}(t, z), \;\; \and \;\; g^{(\ell, d)}(t, z_{\ell}=1,
z_{d}=z) = g^{(d)}(t, z)$ from (\ref{8}) we obtain the following
two equations:
\begin{equation} \label{10}
g^{(\ell)}(t, z)= e^{-Qt}\;z + Q\;\int_{0}^{t}e^{-Q(t-t')}\;
q\left[g^{(\ell)}(t', z)\right]\;dt',
\end{equation}
\begin{equation} \label{11}
g^{(d)}(t, z)= e^{-Qt} + z\;Q\;\int_{0}^{t}e^{-Q(t-t')}\;
q\left[g^{(d)}(t', z)\right]\;dt'.
\end{equation}
It is an elementary task to prove that the integral equation
(\ref{10}) is equivalent to the differential equation
\begin{equation}\label{12}
\frac{\partial g^{(\ell)}(t, z)}{\partial t} = - Q\;g^{(\ell)}(t,
z) + Q\;q\left[g^{(\ell)}(t, z)\right],
\end{equation}
with the initial condition $\lim_{t \downarrow 0}\;g^{(\ell)}(t, z)
= z$, while Eq.(\ref{11}) to
\begin{equation}\label{13}
\frac{\partial g^{(d)}(t, z)}{\partial t} = - Q\;g^{(d)}(t, z) +
z\;Q\;q\left[g^{(d)}(t, z)\right],
\end{equation}
with the initial condition $\lim_{t \downarrow 0}\;g^{(d)}(t, z) =
1$.

Finally, it can be easily shown that $g(t, z) =
\mathbf{E}\{z^{\mu(t)}\}$ satisfies the integral equation
\begin{equation} \label{14}
g(t, z)= e^{-Qt}\;z + z\;Q\;\int_{0}^{t}e^{-Q(t-t')}\;q\left[g(t',
z)\right]\;dt'.
\end{equation}
For the sake of completeness we derive from Eq.~(\ref{14}) the
equivalent differential equation
\[ \frac{\partial g(t, z)}{\partial t} = - Q\;g(t, z),
+ Q z\; q\left[g(t, z)\right] \]  the initial condition of which is
$\lim_{t \downarrow 0}\; g(t, z) = z$.

It can be shown \cite{sevast'yanov71} that each of the equations
for the generating functions has just one regular solution.

\subsection{Joint distribution of the numbers of living and dead
end-nodes}

Now we would like to  determine the joint distribution of
$\chi_{\ell}(t)$ and $\chi_{d}(t)$, i.e., the probability
\[ {\mathcal P}\{\chi_{\ell}(t)=n_{\ell},\; \chi_{d}(t)=n_{d}|{\mathcal S}_{0}\} =
p_{e}^{(\ell,d)}(t, n_{\ell}, n_{d}). \] It is evident that
\[ p_{e}^{(\ell,d)}(t, n_{\ell}, n_{d}) = e^{-Qt} \delta_{n_{\ell},1} \delta_{n_{d},0}
+ \] \[ + Q \int_{0}^{t}
e^{-Q(t-t')}\;\left[f_{0}\delta_{n_{\ell},0} \delta_{n_{d},1} +
\sum_{k=1}^{\infty} f_{k}\;R_{e}^{(\ell,d)}(t', n_{\ell}, n_{d}
\vert k)\right]\;dt', \] where
\[ R_{e}^{(\ell,d)}(t', n_{\ell}, n_{d} \vert k) =
\sum_{n_{\ell,1}+\cdots+n_{\ell,k}=n_{\ell}}\;\sum_{n_{d,1}+\cdots+n_{d,k}=n_{d}}\;
\prod_{j=1}^{k}p_{e}^{(\ell,d)}(t', n_{\ell,j}, n_{d,j}).  \]
Simple calculations show that the generating function
\[ g_{e}^{(\ell,d)}(t, z_{\ell}, z_{d}) = \sum_{n_{\ell}=0}^{\infty}\;
\sum_{n_{d}=0}^{\infty} p_{e}^{(\ell,d)}(t, n_{\ell},
n_{d})\;z_{\ell}^{n_{\ell}}\;z_{d}^{n_{d}} \] satisfies the
equation
\begin{equation} \label{15}
g_{e}^{(\ell,d)}(t, z_{\ell}, z_{d}) = e^{-Qt}z_{\ell} - f_{0}(1 -
z_{d})(1 - e^{-Qt}) + Q \int_{0}^{t}e^{-Q(t-t')}\; q\left[
g_{e}^{(\ell,d)}(t, z_{\ell}, z_{d})\right]\;dt',
\end{equation}
which will be used for the determination of the correlation
between the random variables $\chi_{\ell}(t)$ and $\chi_{d}(t)$.
For the sake of brevity the corresponding differential equation is
omitted.

\subsubsection{Remark}

There is no need to write down the equation for generating
function of $\chi_{\ell}(t)$, since $g_{e}^{(\ell)}(t, z) =
g^{\ell}(t)$ which is the solution of Eq.~(\ref{10}). The equation
for the generating function $g_{e}^{(d)}(t, z)$, however, one can
obtain from (\ref{15}) by substitutions $z_{\ell} = 1$ and $z_{d}
= z$, and so one has
\begin{equation} \label{16}
g_{e}^{(d)}(t, z) = e^{-Qt} - (1 - z) f_{0}(1 - e^{-Qt}) +
Q\;\int_{0}^{t} e^{-Q(t-t')}q\left[g_{e}^{(d)}(t', z)\right]\;dt'.
\end{equation}

In order to gain an insight into the interplay between the living
and dead end-nodes it seems to be useful to calculate the
probability distribution of the random function $\chi(t) =
\chi_{\ell}(t) + \chi_{d}(t)$.  The equation for the generating
function is easily obtained as
\[ g_{e}(t, z) = \sum_{n=0}^{\infty} {\mathcal P}\{\chi(t)=
n|{\mathcal S}_{0}\}\;z^{n} = \sum_{n=0}^{\infty} p_{e}(t,
n)\;z^{n}, \] where $p_{e}(t, n)$ is the probability that the
number of all end-nodes at $t \geq 0$ is equal to $n$ provided
that at $t=0$ the tree was in the state ${\mathcal S}_{0}$.
Applying the usual argumentations one has
\begin{equation} \label{17}
g_{e}(t, z) = e^{-Qt}\;z - (1 - z) f_{0}(1 - e^{-Qt}) +
Q\;\int_{0}^{t} e^{-Q(t-t')}q\left[g_{e}(t', z)\right]\;dt'.
\end{equation}

\section{Average characteristics}

\subsection{Correlations between the numbers of living and dead
nodes}

The study of how the random functions $\mu_{\ell}(t)$ and
$\mu_{d}(t)$ are related involves the analysis of correlation
between them. First we should calculate the covariance function
\begin{equation} \label{18}
{\bf Cov}\{\mu_{\ell}(t) \mu_{d}(t)\} = {\bf E}\{\mu_{\ell}(t)
\mu_{d}(t)\} - {\bf E}\{\mu_{\ell}(t)\}{\bf E}\{\mu_{d}(t)\},
\end{equation}
and then the correlation function
\begin{equation} \label{19}
C^{(\ell,d)}(t, q_{1}, q_{2}) = \frac{{\bf Cov}\{\mu_{\ell}(t)
\mu_{d}(t)\}}{{\bf D}\{\mu_{\ell}(t)\} {\bf D}\{\mu_{d}(t)\}}.
\end{equation}
As is apparent, we must determine the expectations and variances of
$\mu_{\ell}(t)$ and $\mu_{d}(t)$. Taking into account the
relationships
\[ {\bf E}\{\mu_{\ell}(t)\} = \left(\frac{\partial g^{(\ell)}(t,
z)}{\partial z}\right)_{z=1} = m_{1}^{(\ell)}(t) \] and
\[ {\bf E}\{\mu_{d}(t)\} = \left(\frac{\partial g^{(d)}(t,
z)}{\partial z}\right)_{z=1} = m_{1}^{(d)}(t) \] after simple
algebra we find
\begin{equation} \label{20}
m_{1}^{(\ell)}(t) = e^{-\alpha t} \;\;\;\;\;\; \mbox{and}
\;\;\;\;\; m_{1}^{(d)}(t) = \left\{ \begin{array}{ll}
\frac{1}{1-q_{1}}\;\left[1 - e^{-\alpha t}\right], & \mbox{if $q_{1} \neq 1$,} \\
\mbox{ } & \mbox{ } \\
Qt, &  \mbox{if $q_{1} = 1$,}
\end{array} \right.
\end{equation}
where $\alpha = (1-q_{1})Q$. It is here to mention that the tree
evolution is called subcritical, if $q_1 < 1$, critical, if $q_1 =
1$, and supercritical, if $q_1 > 1$.

The formulas
\begin{equation} \label{20a}
\mathbf{D}^{2}\{\mu_{\ell}(t)\} = m_{2}^{(\ell)}(t) +
m_{1}^{(\ell)}(t)\left[1 - m_{1}^{(\ell)}(t)\right]
\end{equation}
and
\begin{equation} \label{20b}
\mathbf{D}^{2}\{\mu_{d}(t)\} = m_{2}^{(d)}(t) +
m_{1}^{(d)}(t)\left[1 - m_{1}^{(d)}(t)\right]
\end{equation}
show that we must determine the second factorial moments
\[ m_{2}^{(\ell)}(t) = \left(\frac{\partial^{2} g^{(\ell)}(t,
z)}{\partial z^{2}}\right)_{z=1}\;\;\;\;\;\; \mbox{and}
\;\;\;\;\;\;  m_{2}^{(d)}(t) = \left(\frac{\partial^{2} g^{(d)}(t,
z)}{\partial z^{2}}\right)_{z=1}. \] After elementary
calculations we find that
\begin{equation} \label{22}
m_{2}^{(\ell)}(t) = \left\{ \begin{array}{ll}
        \frac{q_2}{1 - q_1}\;[1 - e^{-\alpha t}]\;e^{-\alpha t}, &
        \mbox{if $q_1 \not= 1$,} \\
        \mbox{ } & \mbox{ } \\
        q_2\;Qt, & \mbox{if $q_1=1$,}
                   \end{array} \right.
\end{equation}
and
\begin{equation} \label{23}
m_{2}^{(d)}(t) = \left\{ \begin{array}{ll} a_{1}^{(d)}\;[1 -
e^{-\alpha t}] + a_{2}^{(d)}\;Qt\;e^{-\alpha t} +
a_{3}^{(d)}\;e^{-\alpha t}\; [1 - e^{-\alpha t}], &  \mbox{if
$q_{1} \neq 1$,} \\
\mbox{ } & \mbox{ } \\
(Qt)^{2} + \frac{1}{3}\;q_{2}\;(Qt)^{3}, &  \mbox{if $q_{1} = 1$,}
\end{array} \right.
\end{equation}
where
\[ a_{1}^{(d)} = \frac{1}{(1-q_{1})^{2}}\left(2\;q_{1} +
\frac{q_{2}}{1-q_{1}}\right), \;\;\;\; a_{2}^{(d)} =
-\frac{2}{1-q_{1}}\left(q_{1} + \frac{q_{2}}{1-q_{1}}\right), \]
and \[  a_{3}^{(d)} = \frac{q_{2}}{(1-q_{1})^{3}}. \]  By
substituting the first and second factorial moments of
Eq.~(\ref{20a}) for formulas (\ref{20}) and (\ref{22}), we obtain
that
\begin{equation} \label{24a}
{\bf D}^{2}\{\mu_{\ell}(t)\} = \left\{ \begin{array}{ll}
        \frac{{\bf E}\{(\nu-1)^{2}\}}{1 - q_{1}}\;[1 - e^{-\alpha t}]
        \; e^{-\alpha t}, & \mbox{if $q_1 \not= 1$,} \\
        \mbox{ } & \mbox{ } \\
        {\bf D}^{2}\{\nu\}\;Qt, & \mbox{if $q_1=1$,}
                   \end{array} \right.
\end{equation}
where ${\bf E}\{(\nu-1)^{2}\} = (1 - q_{1})^{2} + {\bf
D}^{2}\{\nu\}$.\footnote{In the case of subcritical evolution the
variance has a maximum at $t_{max} = \log 2/\alpha$ which, as seen,
is independent of the form of the distribution of $\nu$.} Applying
similar substitution in (\ref{20b}), if $q_{1} \neq 1$, then  we
arrive at
\begin{equation} \label{25}
{\bf D}^{2}\{\mu_{d}(t)\} = v_{1}^{(d)}\;[1 - e^{-\alpha t}] +
v_{2}^{(d)}\;Qt \;e^{-\alpha t} + v_{3}^{(d)}\;e^{-\alpha t}\; [1
- e^{-\alpha t}],
\end{equation}
where
\[ v_{1}^{(d)} =  \frac{{\bf D}^{2}\{\nu\}}{(1-q_{1})^{3}},
\;\;\;\; v_{2}^{(d)} = - 2\;\frac{{\bf
D}^{2}\{\nu\}}{(1-q_{1})^{2}},  \;\;\;\; v_{3}^{(d)} = \frac{{\bf
E}\{(\nu-1)^{2}\}}{(1-q_{1})^{3}}. \] If $q_{1} = 1$, i.e., the
evolution is critical, then
\begin{equation} \label{26}
{\bf D}^{2}\{\mu_{d}(t)\}  =  Qt + \frac{1}{3}\;q_{2}\;(Qt)^{3}.
\end{equation}
One can see that the limit values of the variance of the number of
dead nodes are given by the formula
\[ \lim_{t \rightarrow \infty}{\bf D}^{2}\{\mu_{d}(t)\} =  \left\{ \begin{array}{ll}
\frac{{\bf D}^{2}\{\nu\}}{(1-q_{1})^{3}}, & \mbox{if $q_{1}<1$,} \\
            \mbox{} & \mbox{} \\
              \infty, &  \mbox{if $q_{1} \geq 1$.}
             \end{array} \right. \]

In order to calculate the covariance function we should solve also
the equation determining the mixed moment
\begin{equation} \label{27}
{\bf E}\{\mu_{\ell}(t) \mu_{d}(t)\} = \left[\frac{\partial^{2}
g^{(\ell,d)}(t, z_{\ell}, z_{d})}{\partial z_{\ell} \partial
z_{d}}\right]_{z_{\ell}=z_{d}=1} = m_{1,1}^{(\ell,d)}(t).
\end{equation}
The equation can be written in the form
\[ m_{1,1}^{(\ell,d)}(t) = Q\;q_{1}\;\int_{0}^{t}
e^{-Q(t-t')}\;m_{1,1}^{(\ell,d)}(t')\;dt' + \]
\begin{equation} \label{28}
Q\;\int_{0}^{t} e^{-Q(t-t')}\left[q_{1}\;m_{1}^{(\ell)}(t') +
q_{2}\;m_{1}^{(\ell)}(t')\;m_{1}^{(d)}(t')\right]\;dt'.
\end{equation}
By using  Eq.~(\ref{20}) for $m_{1}^{(\ell)}(t)$ and
$m_{1}^{(d)}(t)$, and introducing the notation $\alpha =
(1-q_{1})\;Q$, it can be proved that
\[ m_{1,1}^{(\ell,d)}(t) = \left\{ \begin{array} {ll}
\frac{{\bf D}^{2}\{\nu\}}{1-q_{1}}\;Qt\;e^{-\alpha t} -
q_{2}\;\frac{1}{(1-q_{1})^{2}}\;e^{- \alpha t}\;(1 - e^{- \alpha
t}), & \mbox{if $q_{1} \neq 1$,} \\
\mbox{ } &  \mbox{ } \\
Qt + \frac{1}{2} {\bf D}^{2}\{\nu\} (Q t)^{2}, & \mbox{if $q_{1} =
1$} \end{array} \right. \]  is the unique solution of the integral
equation (\ref{26}). After a brief calculation we have
\begin{equation} \label{29}
{\bf Cov}\{\mu_{\ell}(t) \mu_{d}(t)\} = \left\{ \begin{array} {ll}
\frac{{\bf D}^{2}\{\nu\}}{1-q_{1}}\;Qt\;e^{-\alpha t} - \left[1 +
\frac{{\bf D}^{2}\{\nu\}}{(1-q_{1})^{2}}\right]\;e^{- \alpha
t}\;(1 - e^{- \alpha t}), & \mbox{if $q_{1} \neq 1$,} \\
\mbox{ } & \mbox{ } \\
\frac{1}{2}\;{\bf D}^{2}\{\nu\}\;(Qt)^{2}, & \mbox{if $q_{1} =
1$.}
\end{array} \right.
\end{equation}
\begin{figure} [ht]
\protect \centering{
\includegraphics[height=8cm, width=12cm]{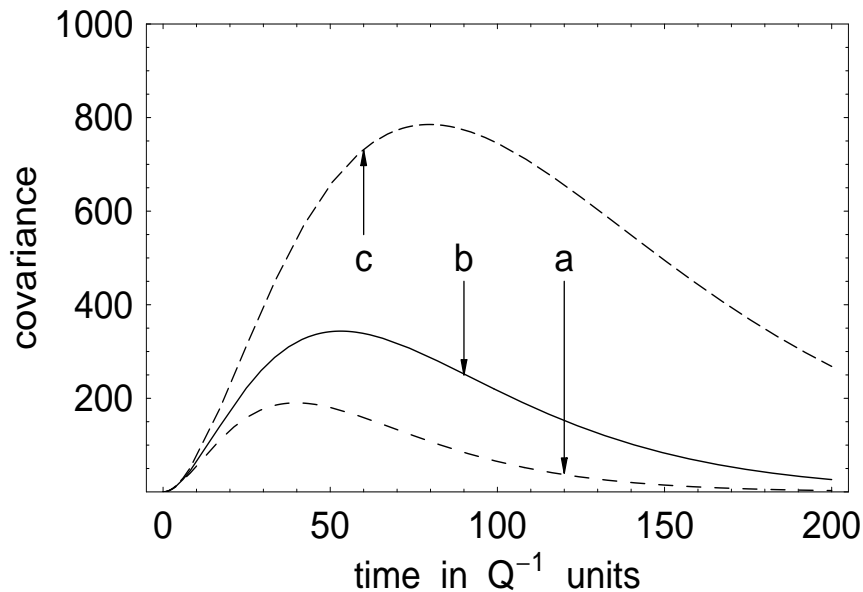}}\protect
\vskip 0.2cm \protect \caption{\label{fig3}\footnotesize{Three
curves illustrating the time dependence of the covariance ${\bf
Cov}\{\mu_{\ell}(t) \mu_{d}(t)\}$ in the case of subcritical
random evolution with the assumption that $\nu$ is of geometric
distribution. The curves {\bf a, b}, and {\bf c} correspond to the
values of $q_{1}=0.96, 0.97$ and $0.98$, respectively.}}
\end{figure}

The dependence of the covariance function ${\bf Cov}\{\mu_{\ell}(t)
\mu_{d}(t)\}$ on the time parameter $Qt$ at three different values
of $q_{1}$ has been calculated by assuming that $\nu$ is of
geometric distribution. The three curves {\bf a, b}, and {\bf c}
corresponding to the values $q_{1}=0.96, 0.97$, and $0.98$,
respectively, are plotted in Fig. \ref{fig3}. One may observe that
each of the covariance curves vs time has a well defined maximum.
The time parameter which belongs to the maximum of the covariance
function depends rather sensitively on $q_{1}$.

Now, we would like to return to the study of the correlation
function defined by (\ref{19}). It is evident that the numbers of
living and dead nodes are correlated, but it is worthwhile to note
that the time dependence of the correlation function
$C^{(\ell,d)}(t, q_{1}, q_{2})$ has some specific features, which
follow from properties of the covariance function and the
variances of $\mu_{\ell}(t)$ and $\mu_{d}(t)$. For example, one
can show that in the case of subcritical evolution the correlation
function, after a sharp increase, converges to zero when $Qt
\rightarrow \infty$, while in the case of supercritical evolution
it rapidly reaches the value $1$.
\begin{figure} [ht]
\protect \centering{
\includegraphics[height=8cm, width=12cm]{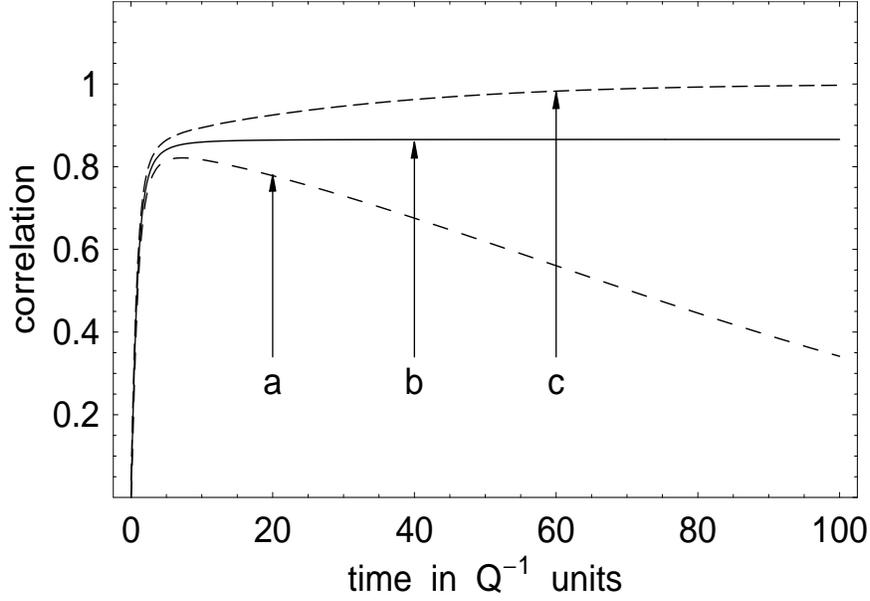}}\protect
\vskip 0.2cm \protect \caption{\label{fig5}\footnotesize{Time
dependence of the correlation between the numbers living and dead
nodes if $\nu$ is of geometric distribution. The curves {\bf a,
b,} and {\bf c} belong to $q_{1}$ values $0.95, \; 1,$ and $1.05$,
respectively.}}
\end{figure}
This behavior is illustrated in Fig. \ref{fig5} for the case of
$\nu$ following a  geometric distribution. The curves {\bf a, b,}
and {\bf c} belong to values $0.95, \; 1,$ and $1.05$ of $q_{1}$,
respectively. It seems to be interesting to present some curves
reflecting the dependence of $C^{(\ell,d)}(t, q_{1}, q_{2})$ on
$q_{1}$ at different time instants.
\begin{figure} [ht!]
\protect \centering{
\includegraphics[height=8cm, width=12cm]{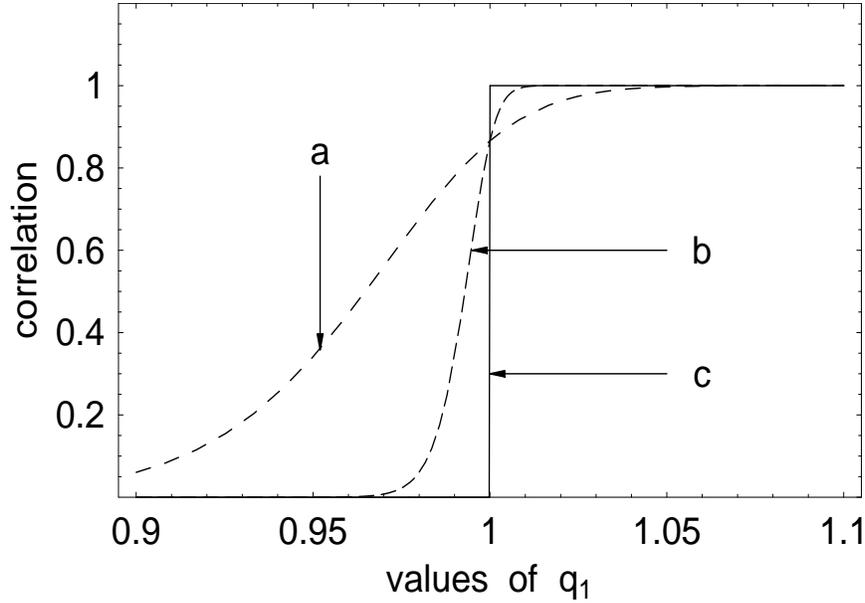}}\protect
\vskip 0.2cm \protect
\caption{\label{fig9}\footnotesize{Dependence of the correlation
between the numbers of living and dead nodes on $q_{1}$ at time
instants $Qt=100$, ({\bf a}), $Qt=500$, ({\bf b}) and $Qt=\infty$,
({\bf c}).}}
\end{figure}
In Fig. \ref{fig9} we can see that the transition of the correlation
function from $0$ to $1$ becomes sharper and sharper as $Qt$
increases. The curves {\bf a} and {\bf b} belong to the time
parameters $Qt=100$ and $Qt=500$, respectively, while curve {\bf c}
shows how  the correlation function depends on $q_{1}$ at $Qt =
\infty$.

The most surprising result can be obtained when the random evolution
is critical, i.e., $q_{1}=1$.  Then we have
\begin{equation} \label{30}
C^{(\ell,d)}(t, 1, q_{2}) = \frac{\sqrt{3}}{2}\;\left[1 +
\frac{3}{{\bf D}^{2}\{\nu\}\;(Qt)^{2}}\right]^{-1/2},
\end{equation}
which shows that the limit value of the correlation
\begin{equation} \label{31}
\lim_{Qt \rightarrow \infty}\; C^{(\ell,d)}(t, 1, q_{2}) =
\frac{\sqrt{3}}{2}
\end{equation}
is independent of the distributions of $\nu$ and $\tau$. This very
important limit law expresses the fact that the relation between the
numbers of living and dead nodes near the critical state, and at
time moments sufficiently far from the beginning of the process,
becomes almost independent of rules controlling the evolution. This
independence can be attributed to the appearance of the universal
properties of the evolution.

\subsection{Correlations between the numbers of living and dead
end-nodes}

The dynamics of the random tree evolution is determined by the
end-nodes. It is evident that living end-nodes are responsible for
the development of a tree, while dead end-nodes represent points
where the tree is not capable of any variation. In order to have an
insight into the dynamics of the tree evolution it is useful to
study the correlation function of $\chi_{\ell}(t)$ and
$\chi_{d}(t)$. First we calculate the covariance function
\begin{equation} \label{32}
{\bf Cov}\{\chi_{\ell}(t) \chi_{d}(t)\} = {\bf E}\{\chi_{\ell}(t)
\chi_{d}(t)\} - {\bf E}\{\chi_{\ell}(t)\}{\bf E}\{\chi_{d}(t)\},
\end{equation}
and then the correlation function
\begin{equation} \label{33}
C_{e}^{(\ell,d)}(t, q_{1}, q_{2}) = \frac{{\bf
Cov}\{\chi_{\ell}(t) \chi_{d}(t)\}}{{\bf D}\{\chi_{\ell}(t)\} {\bf
D}\{\chi_{d}(t)\}}.
\end{equation}
Since $\chi_{\ell}(t) = \mu_{\ell}(t)$, we have to deal only with
the determination of the variance $\mathbf{D}^{2}\{\chi_{d}\}$.
Omitting the details of calculations,  if $q_{1} \neq 1$, we have
\[ \mathbf{D}^{2}\{\chi_{d}(t)\} =
\frac{f_{0}^{2}}{(1-q_{1})^{2}}\left[\left(\frac{q_{2}}{1-q_{1}}+\frac{1-q_{1}}{f_{0}}
- 1\right)\;\left(1 - e^{-\alpha t}\right) - 2 q_{2}
Qt\;e^{-\alpha t} + \right. \]
\begin{equation} \label{34}
\left. + \left(\frac{q_{2}}{1-q_{1}}+1\right)\;\left(1 -
e^{-\alpha t}\right)\;e^{-\alpha t}\right],
\end{equation}
while, if $q_{1} = 1$, we obtain the following simple formula:
\begin{equation} \label{35}
\mathbf{D}^{2}\{\chi_{d}(t)\} = f_{0} Qt \left(1 + \frac{1}{3}
q_{2} (qt)^{2}\right).
\end{equation}
It is worthwhile to remark that if the tree evolution is
subcritical, then and only then does the variance of the number of
dead end-nodes converge to a finite limit value given by
\[ \lim_{t \rightarrow \infty}\;\mathbf{D}^{2}\{\chi_{d}(t)\} =
\frac{f_{0}^{2}}{(1-q_{1})^{2}}\left(\frac{q_{2}}{1-q_{1}}+\frac{1-q_{1}}{f_{0}}
- 1\right). \]

As regards the covariance function from the generating function
$g_{e}^{(\ell,d)}(t, z_{\ell}, z_{d}) $ defined by (\ref{15}) we
should derive the second mixed moment
\[ m_{1,1}^{(\ell,d)}(t|e) =
\left(\frac{\partial^{2}g_{e}^{(\ell,d)}(t, z_{\ell}, z_{d})}
{\partial z_{\ell}\;\partial z_{d}}\right)_{z_{\ell}=z_{d}=1} \]
which is determined by the following integral equation:
\begin{equation} \label{36}
m_{1,1}^{(\ell,d)}(t|e) = q_{1} Q \int_{0}^{t}
e^{-Q(t-t')}\;m_{1,1}^{(\ell,d)}(t'|e)\;dt' + q_{2} Q \int_{0}^{t}
e^{-Q(t-t')}\;m_{1}^{(\ell)}(t'|e)\;m_{1}^{(d)}(t'|e)\;dt'.
\end{equation}
Since
\[ m_{1}^{(\ell)}(t|e) = m_{1}^{(\ell)}(t) = e^{-\alpha
t}\;\;\;\;\;\; \mbox{and} \;\;\;\;\; m_{1}^{(d)}(t) =
f_{0}\;\frac{1-e^{-\alpha t}}{1-q_{1}}, \] we obtain the solution
in the form:
\begin{equation} \label{37}
\mathbf{Cov}\{\chi_{\ell}(t) \chi_{d}(t)\} = \left\{
\begin{array}{ll} \frac{f_{0}}{1-q_{1}}\left[q_{2} Qt -
\left(\frac{q_{2}}{1-q_{1}}+1\right)(1-e^{-\alpha
t})\right]\;e^{-\alpha t}, & \mbox{if $q_{1} \neq 1$,} \\
\mbox{ } & \mbox{ } \\
f_{0} Qt \left(\frac{1}{2} q_{2} Qt - 1\right),& \mbox{if $q_{1} =
1$.} \end{array} \right.
\end{equation}

\begin{figure} [ht!]
\protect \centering{
\includegraphics[height=4.8cm, width=8cm]{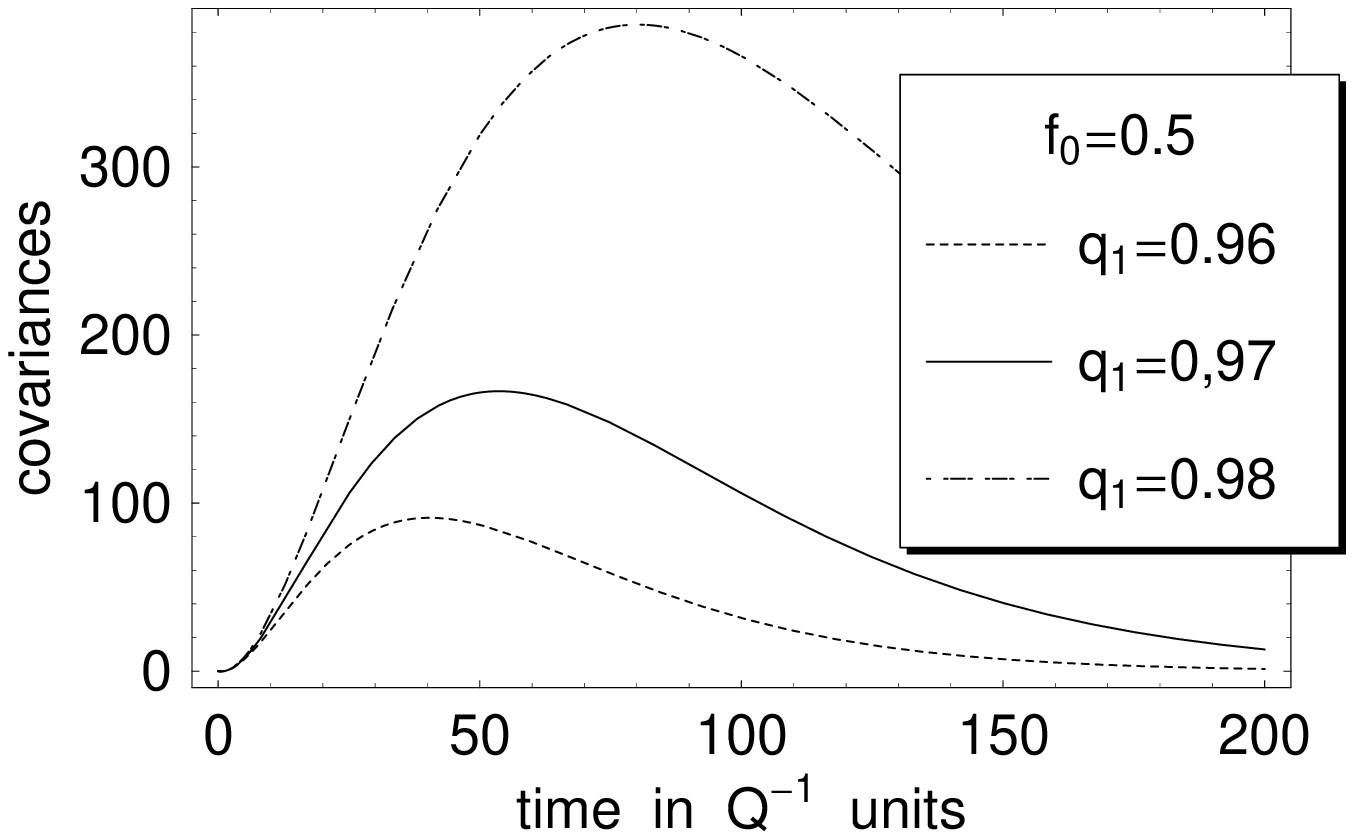}}\protect
\protect \centering{
\includegraphics[height=4.8cm, width=8cm]{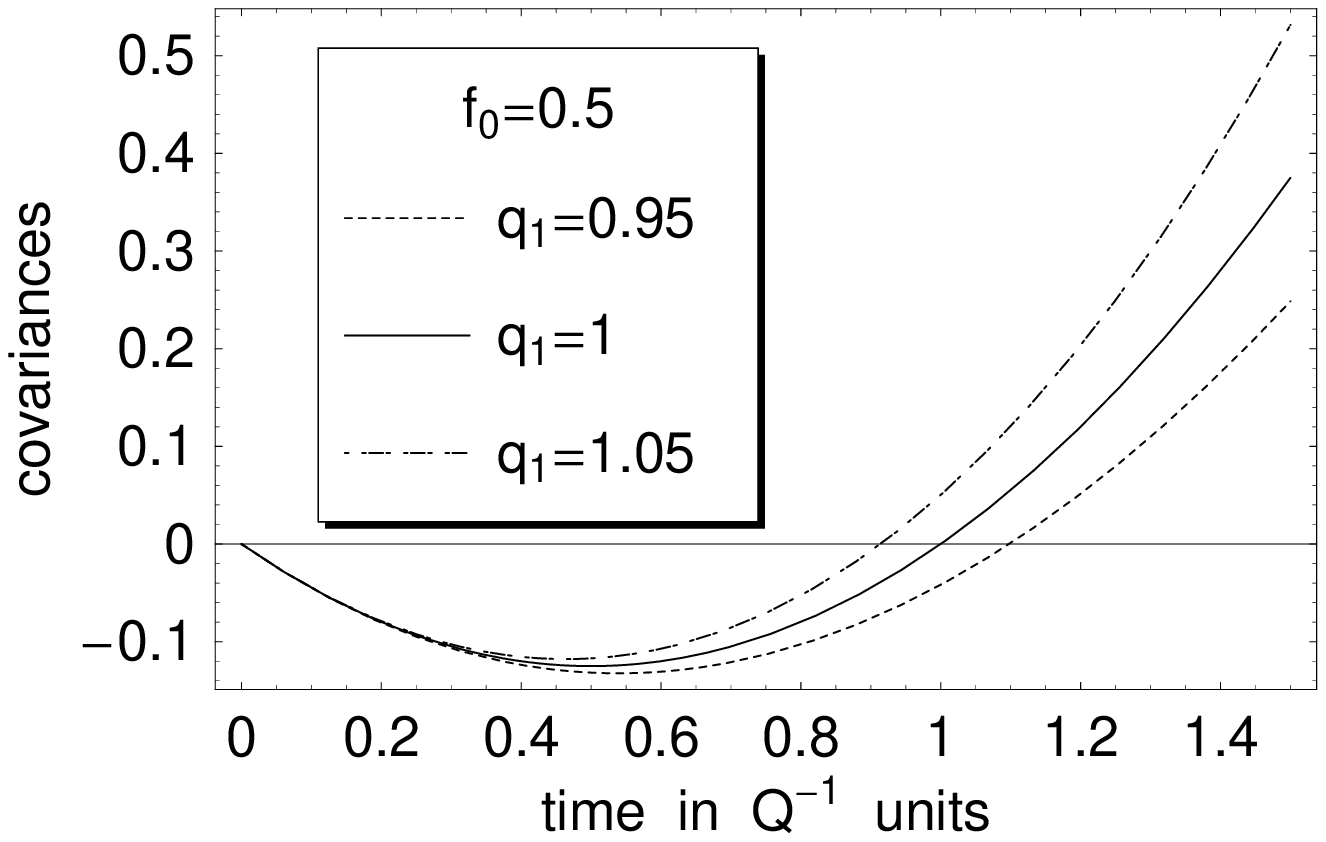}}\protect
\vskip 0.2cm \protect
\caption{\label{figa}\footnotesize{Dependence of the covariance of
the numbers of living and dead end-nodes on $Qt$ at different
values of $q_{1}$.}}
\end{figure}

As is to be seen in Fig. \ref{figa}, the covariance of the numbers
of living and dead end-nodes changes its sign at the very beginning
of the evolution. The negative covariance just after the start is a
consequence of the simple fact that the root of the tree can be but
a living end-node at the time instant $t=0$. The calculations refer
to the case when $\nu$ is of geometric distribution; this however,
does not affect the generality of the conclusion.

\begin{figure} [ht!]
\protect \centering{
\includegraphics[height=8cm, width=12cm]{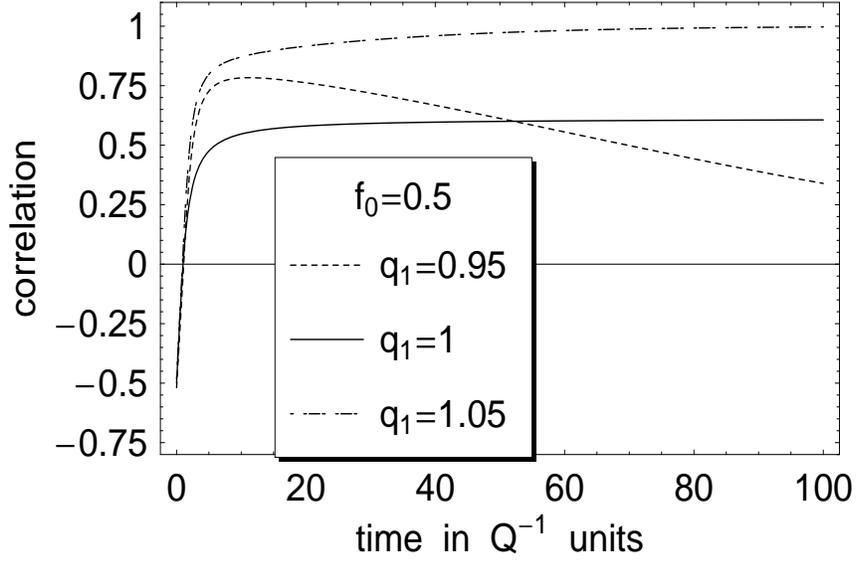}}\protect
\vskip 0.2cm \caption{\label{figb}\footnotesize{Dependence of the
correlation between the numbers of living and dead end-nodes on
$Qt$ at different values of $q_{1}$ .}}
\end{figure}

For the sake of completeness we calculated the time dependence of
the correlation function of the numbers of living and dead end-nodes
at values of $q_{1}=0.95, 1, 1.05$ assuming again that $\nu$ is of
geometric distribution. As expected, Fig. \ref{figb} shows
convincingly the change realizing in the character of the
correlation near the beginning of the evolution. There is no need to
write down the lengthy expression of the correlation function but it
seems to be useful to present it when the tree evolution is
critical. In this case one obtains
\begin{equation} \label{38}
C_{e}^{(\ell,d)}(t) = \frac{1}{2}\;\sqrt{\frac{3
f_{0}}{q_{2}}}\;\frac{q_{2} Qt - 2}{\sqrt{3 + q_{2} (Qt)^{2}}}.
\end{equation}
It is interesting to note that the correlation in critical evolution
reaches its asymptotic value $\sqrt{3 f_{0}}/2$ very soon. (If
$f_{0}=0.5$, then $C_{e}^{(\ell,d)}(\infty) \approx 0.612$ and this
is what is seen in Fig. \ref{figb}.)

\section{Survival probability}

\subsection{General considerations}

It is obvious that the evolution of a random tree will stop at the
time instant $\theta$ which satisfies the equation
$\mu_{\ell}(\theta)=0$ with probability 1. The random variable
$\theta$ is called the lifetime of the tree. In order to determine
its distribution function
\begin{equation} \label{39}
{\mathcal P}\{\theta \leq t|{\mathcal S}_{0}\}= L(t),
\end{equation}
one must recognize that the probability ${\mathcal
P}\{\mu_{\ell}(t)=0|{\mathcal S}_{0}\} = p^{(\ell)}(t, 0)$ to find
zero living node at the time moment $t \geq 0$ in a tree is the same
as the probability that the lifetime $\theta$ of that tree is not
larger than $t \geq 0$, therefore, one can write ${\mathcal
P}\{\theta \leq t|{\mathcal S}_{0}\} = {\mathcal
P}\{\mu_{\ell}(t)=0| {\mathcal S}_{0}\}$, i.e.,
\begin{equation} \label{40}
L(t) = p^{(\ell)}(t,0) = \lim_{z \downarrow 0}\; g^{(\ell)}(t, z).
\end{equation}
It is clear that if $0 < t_{1} \leq t_{2}$ then $L(t_{1}) \leq
L(t_{2})$, i.e., $L(t)$ is a nondecreasing function of its argument,
hence the limit relation
\begin{equation} \label{41}
\max_{0<t\leq \infty}\;L(t) = \lim_{t \rightarrow \infty}\; L(t) =
L_{\infty} \leq 1
\end{equation}
must hold. We will call the quantity $L_{\infty}$ extinction
probability, and will now prove the following statement: if $q_{1}
\leq 1$, i.e., the random evolution is not supercritical, then
$L_{\infty}=1$, while if $q_{1} > 1$, i.e., the evolution is
supercritical, then $L_{\infty}$ is equal to the non-negative,
single root of the function $\psi(y) = q(y) - y, \;\; y \in [0, 1]$.
(It is evident that $\psi(1) = 0$.)

For the proof we exploit the fundamental property of the
generating function $g^{(\ell)}(t, z)$ which is expressed by the
equation $g^{(\ell)}(t+u, z) = g^{(\ell)}\left[t, g^{(\ell)}(u,
z)\right]$. Applying the relation (\ref{41}) we have $L(t+u) =
g^{(\ell)}\left[t, L(u)\right]$ and since
\[  \lim_{u \rightarrow \infty}\;L(t+u)=
\lim_{u \rightarrow \infty}\;L(u) = L_{\infty}, \] we can write
for every $t \geq 0$ that $L_{\infty} = g^{(\ell)}(t,
L_{\infty})$. Plugging $g^{(\ell)}(t, L_{\infty})$ into the
equation
\begin{equation} \label{42}
\frac{\partial g^{(\ell)}(t,z)}{\partial} =
Q\;q\left[g^{(\ell)}(t,z)\right] - Q\; g^{(\ell)}(t,z),
\end{equation}
we obtain
\begin{equation} \label{43}
q\left(L_{\infty}\right) - L_{\infty} = 0.
\end{equation}
Considering that $q(y)$ is a probability generating function,
i.e., $\lim_{y \uparrow 1}\;q(y) = 1$, then according to a
well-known theorem of generating functions, it is clear that if
$q_{1} > 1$ then Eq. (\ref{43}) besides the trivial fixed-point
$1$ must have also another, smaller than $1$, non-negative
fixed-point $L_{\infty}$, and this is what we wanted to prove.

Now, we would like to define the survival probability. Obviously,
\begin{equation} \label{44}
S(x) = 1 - L(x),
\end{equation}
is the probability to find the tree at the time instant $x=Qt$ in
a living state, and therefore, $S(x)$ can be called  the survival
probability. By taking into account the properties of $L(x)$ one
obtains
\begin{equation} \label{45}
\lim_{Qt \rightarrow \infty} S(x) = \left\{ \begin{array}{ll}
         0, &  \mbox{if $q_{1} \leq 1$,} \\
         \mbox{} & \mbox{} \\
         S_{\infty}=1-L_{\infty}, & \mbox{if $q_{1} > 1$.}
         \end{array} \right.
\end{equation}
Clearly, from  Eqs. (\ref{42}), (\ref{40}), and (\ref{44}) one can
derive the equation
\begin{equation} \label{46}
\frac{dS}{dx} = -q(1-S) - S + 1,
\end{equation}
which has the solution
\begin{equation} \label{47}
x(S) = \int_{S(x)}^{1} \frac{dy}{q(1-y) + y - 1},
\end{equation}
at the initial condition $S(0)=1$.

\subsection{Calculations in the case of a quadratic $q(z)$
function}

In order to demonstrate the characteristic features of the
lifetime of random trees we will determine the survival
probability in the case when the generating function of the
offspring number $\nu$ is a quadratic expression defined by
(\ref{5a}). By using Eq. (\ref{46})  we obtain
\begin{equation} \label{48}
\frac{dS}{dx} = -(1-q_{1})\;S  - \frac{1}{2}\;q_{2}\;S^{2},
\end{equation}
and taking into account the initial condition $S(0)=1$ we have the
solution in the form
\begin{equation} \label{49}
S(x) = \left\{ \begin{array}{ll} e^{-(1-q_{1})x} \left[1 +
\frac{q_{2}}{2(1-q_{1})}\;(1 -
e^{-(1-q_{1})x})\right]^{-1}, & \mbox{if $q_{1}<1$,} \\
\mbox{} & \mbox{} \\
\frac{2}{2 + q_{2}\;x}, & \mbox{if $q_{1}=1$,} \\
\mbox{} & \mbox{} \\
2\;\frac{q_{1}-1}{q_{2}}\;\left[1 + (1 -
2\;\frac{q_{1}-1}{q_{2}})\;e^{-(q_{1}-1)x}\right]^{-1}, & \mbox{if
$q_{1}>1$}. \end{array} \right.
\end{equation}

\begin{figure} [ht!]
\protect \centering{
\includegraphics[height=4.8cm, width=8cm]{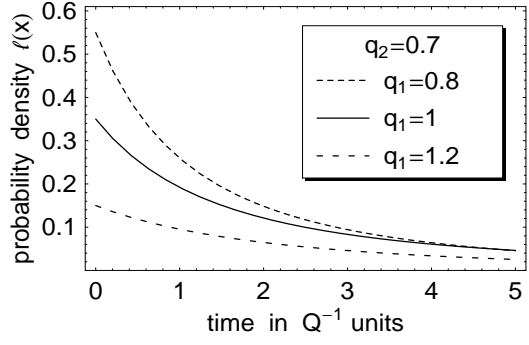}}\protect
\vskip 0.2cm \caption{\label{figc}\footnotesize{Dependence of the
probability density function of the lifetime on $x=Qt$. }}
\end{figure}

The density function of the lifetime measured in units of $Q^{-1}$
can be calculated by using the relation
\[ \ell(x) = \frac{dL(x)}{dx} = - \frac{dS(x)}{dx}. \]
It is elementary to show that the density function is decreasing
monotonously from $\ell(0) = f_{0}$ to zero. In Fig. \ref{figc} one
can see the density function curves versus time $x=Qt$ for
subcritical, critical, and supercritical trees.

From (\ref{49}) one can see that
\begin{equation} \label{50}
\lim_{x=Qt \rightarrow \infty} S(x) = \left\{ \begin{array}{ll}
         0, &  \mbox{if $q_{1} \leq 1$,} \\
         \mbox{} & \mbox{} \\
         S_{\infty}=2\;\frac{q_{1}-1}{q_{2}}, & \mbox{if $q_{1} > 1$.}
         \end{array} \right.
\end{equation}
For more detailed analysis of asymptotic values for $S(x)$ see Ref.
\cite{lpal02}.

Finally, in this section let us calculate the expectation and the
variance of the tree lifetime $\theta$ in units of $Q^{-1}$. For
this purpose it seems to be useful to determine the characteristic
function of the random variable $Q \theta$. Since the moments ${\bf
E}\{(Q\theta)^{j}\}, \;\; j=1,2,\ldots\;$ do not exist if $q_{1}
\geq 1$, the calculations are restricted to the case when $q_{1} <
1$. One can write
\[ \varphi(\omega) = {\bf E}\{e^{-\omega Q\theta}\} =
\int_{0}^{\infty} e^{-\omega x}\;dL(x) = \int_{0}^{1} e^{-\omega
x(y)}\;dy, \] where $\omega$ is a complex number with $Re \omega
\geq 0$. Performing the substitution
\[ x(y) = - \frac{1}{1-q_{1}}\;\ln\; y\;\frac{1 +
2\frac{1-q_{1}}{q_{2}}}{y + 2\frac{1-q_{1}}{q_{2}}} \]  one
obtains the characteristic function
\begin{equation} \label{51}
\varphi(\omega) = (1 + \gamma)^{\omega \beta}\;\int_{0}^{1}
\left[\frac{y} {y + \gamma}\right]^{\omega \beta}\;dy,
\end{equation}
where \[ \beta = (1-q_{1})^{-1}\;\;\;\;\;\; \mbox{and}
\;\;\;\;\;\; \gamma = 2\frac{1-q_{1}}{q_{2}}, \;\;\;\;\;\; q_{1} <
1. \]

From the characteristic functions $\varphi(\omega)$ both the
expectation and the variance of the lifetime can be easily
calculated when the tree evolution is subcritical. For the
expectation value one obtains
\begin{equation} \label{52}
{\bf E}\{Q\theta\} = -
\left(\frac{d\varphi(\omega)}{d\omega}\right)_{\omega=0} =
\frac{2}{q_{2}}\;\ln\;\left(1 +
\frac{1}{2}\;\frac{q_{2}}{1-q_{1}}\right).
\end{equation}
If $q_{1} \rightarrow 1$, then the expectation value diverges as
$\ln\;(1-q_{1})^{-1}$.

For the calculation of the variance we need the second moment of
$Q\theta$, which can be immediately obtained from the characteristic
function. Omitting the details of the calculation we may write
\[ {\bf D}^{2}\{Q\theta\} = \]
\begin{equation} \label{53}
= - \left(\frac{2}{q_{2}}\right)^{2}\;\left(1 +
\frac{1}{2}\;\frac{q_{2}}{1-q_{1}}\right)\;\left[\ln\;\left(1 +
\frac{1}{2}\;\frac{q_{2}}{1-q_{1}}\right)\right]^{2} -
\frac{4}{q_{2}\;(1-q_{1})}\;Li_{2}\left(-\frac{1}{2}\frac{q_{2}}{1-q_{1}}\right),
\end{equation}
where $Li_{2}(u) = \sum_{k=1}^{\infty} u^{k}/k^{2}\;\;$ is the so
called Jonqui\`ere's function. When $q_{1}$ is approaching $1$
from below, the fluctuation of the tree lifetime becomes
unlimitedly large, and so in the vicinity of the critical
evolution the average lifetime loses almost completely its
information content.

\section{Conclusions}

Let us summarize the main results of the paper. By introducing the
notions of living and dead nodes, a model of random tree evolution
with continuous time parameter has been constructed. The model
describes the spreading in time of abstract objects which
correspond to nodes. It is to mention that the process of the
random tree evolution analyzed in this paper belongs to the family
of the age-dependent branching processes.

In order to characterize the evolution process, essentially two
basic random functions $\mu_{\ell}(t)$ and $\mu_{d}(t)$ have been
used. The first one is the momentary number of living nodes, while
the second one is that of dead nodes at the time instant $t \geq 0$.
By the assumption that the evolution process controlled by the
lifetime $\tau$ and the offspring number $\nu$ of living nodes,
exact equations have been derived for the generating functions of
the probabilities \[ {\mathcal P}\{\mu_{\ell}(t)=n_{\ell},
\mu_{d}(t)=n_{d} \vert {\mathcal S}_{0}\}, \;\;\;\;\;\; \mbox{and}
\;\;\;\;\;\; {\mathcal P}\{\mu(t) = \mu_{\ell}(t)+\mu_{d}(t) = n
\vert {\mathcal S}_{0}\}, \] where ${\mathcal S}_{0}$ indicates that
at $t = 0$ the tree consisted of a single living node only. To
complete the description of the tree dynamics the notion of
end-nodes has also been defined. It is remarkable that the average
lifetime of living nodes has a role in scaling the time in the
generating function equations, only.

The time dependence of the expectation value of the number of living
nodes has been analyzed, and as expected, there exist three
completely different types of evolution depending on the average
number $q_{1}$ of offspring produced by a single living node. If
$q_{1}<1$, then the evolution is subcritical, if $q_{1} = 1$, then
it is critical, while in the case of $q_{1}>1$ is supercritical.

A specific property of the tree evolution has been discovered,
namely, it has been proved that the correlation between the
numbers of living and dead nodes decreases to zero in subcritical
and increases to $1$ in supercritical evolution, if the time
parameter $Qt$ tends to infinity, but in the case of exactly
critical evolution it converges to a fixed value $\sqrt{3}/2$
which is free of the process parameters.

The stochastic properties of the end-nodes have also been
analyzed, and it has been shown that the correlation between the
numbers of living and dead end-nodes changes its character rather
suddenly at the very beginning of the evolution process. It has
been proved that the correlation function is negative just after
the start of the process, but it becomes positive very sharply
after elapsing a relatively short time. This behavior is the
consequence of the fact that the root of the tree can be but a
living end-node at the time moment $t = 0$.

For the sake of better understanding of the evolution the survival
probability of random trees has been investigated, and exact
expressions have been derived for this probability in the cases of
subcritical, critical, and supercritical evolutions when the
generating function of the offspring number $\nu$ was quadratic.


\begin{thebibliography}{21}

\bibitem{barabasi01} R. Albert and A.-L. Barab\'asi, Rev. Mod. Phys.
{\bf 74}, 47 (2002).

\bibitem{dorogovtsev01} S.N. Dorogovtsev and J.F.F. Mendes,
Adv. Phys. {\bf 51}, 1079 (2002).

\bibitem{dorogovtsev02} S.N. Dorogovtsev and J.F.F. Mendes,
\textit{Evolution of Networks}, (Oxford University Press, New
York, 2003) p. 241.

\bibitem{harris63} T.E. Harris, \textit{The theory of Branching
Processes}, (Springer-Verlag, Berlin-G\"ot\-tin\-gen-Heidel\-berg,
1963) p. 33; for the age-dependent processes see pp. 121-163.

\bibitem{lyons02} R. Lyons and Y. Peres, \textit{Probability on Trees
and Networks}, (Cambridge University Press, Cambridge), (to be
published).

\bibitem{janossy52} L. J\'anossy, Acta Phys. Acad. Sci. Hung.
{\bf 2}, 289 (1952).

\bibitem{reid54} A.T. Bharucha-Reid, Phys. Rev. {\bf 96}, 751
(1954).

\bibitem{lpal58} L. P\'al, Nuovo Cimento, Suppl. {\bf 7}, 25 (1958).

\bibitem{kendall60} D.G. Kendall, Biometrika, {\bf 47}, 13 (1960).

\bibitem{bell63} G.I. Bell, Ann. Phys. (N.Y.) {\bf 21}, 243 (1963)

\bibitem{pazsit99} I. P\'azsit, Phys. Scr. {\bf 59}, 344
(1999).

\bibitem{williams04} M.M.R. Williams, Ann. Nucl. Energy,
{\bf 31}, 933 (2004).

\bibitem{lpal03} L. P\'al, Prog. Nucl. Energy,
{\bf 43}, 5 (2003).

\bibitem{lpal02} L. P\'al, cond-mat/0205650 (unpublished); cond-mat/0211092
(unpublished); cond-mat/0306540) (unpublished).

\bibitem{sevast'yanov71} B.A. Sevast'yanov, \textit{Verzweigungsprozesse},
(Akademie-Verlag, Berlin, 1974) p. 28; for the age-dependent
processes see pp. 229-309.

\end{thebibliography}
\end{document}